\documentclass[twocolumn,aps,prb,reprint]{revtex4-1}
\usepackage[T1]{fontenc}
\usepackage[utf8]{inputenc}
\usepackage{lmodern}
\usepackage{xcolor}
\usepackage{graphicx}

\usepackage{amsmath}
\usepackage{booktabs}
\usepackage{xcolor}
\usepackage{epstopdf}
\usepackage{ulem}
\usepackage{hyperref}
\usepackage{amsfonts}
\usepackage{amssymb}
\begin{document}

\def\Vhrulefill{\leavevmode\leaders\hrule height 0.7ex depth \dimexpr0.4pt-0.7ex\hfill\kern0pt}
\renewcommand{\vec}[1]{\mathbf{#1}}
\newcommand{\ii}{\mathrm{i}}

\title{From chaos to many-body localization: some introductory notes. }

\author{L. Chotorlishvili$^1$, S. Stagraczy\'nski$^2$, M. Sch\"uler$^3$, J. Berakdar$^1$}
\affiliation{$^1$Institut f\"ur Physik, Martin-Luther-Universit\"at Halle-Wittenberg, 06099 Halle, Germany \\ $^2$ Department of Physics and Medical Engineering, Rzeszow University of Technology, al. Powstancow Warszawy 6, 35-959 Rzeszow, Poland\\ $^3$ Department of Physics, University of Fribourg, 1700 Fribourg, Switzerland}
\date{\today}
\begin{abstract}
Staring from the kicked rotator as a paradigm for a system exhibiting classical chaos, we discuss the role of quantum coherence resulting in dynamical localization in the kicked quantum rotator.  In this context, the disorder-induced Anderson localization is also discussed.  Localization in interacting, quantum many-body systems (many-body localization) may also occur in the absence of disorder, and a practical way to identify its occurrence is demonstrated for an interacting spin chain.
\end{abstract}

\maketitle

\section{Hamiltonian Chaos vs. random impurities}
\begin{figure}[ht]
	\includegraphics[width=0.8\columnwidth]{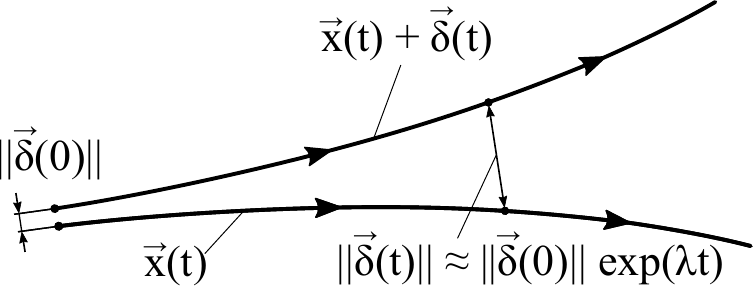}
	\caption{Time evolution of two different phase trajectories.}
	\label{fig:scheme}
\end{figure}

As put by  Edward Lorenz, classical deterministic chaos is \cite{Lorenz}:
{"when the present determines the future, but the approximate present does not approximately determine the future"}.
The instability of phase trajectories is quantified  by the  Lyapunov exponent \cite{Zaslavsky}: In  phase space, the state of the $N$ dimensional dynamical system with $2N$ degrees of freedom $\vec{x}\equiv(\vec{p}_{n},\vec{q}_{n}),~n=1,..N$ is mapped to a single dot and the evolution of the system is described by a phase trajectory circumscribed by the system's state. Let us track the time evolution of two different phase trajectories see Fig.(\ref{fig:scheme}) $\vec{x}(t)+\vec{\delta}(t),~~\vec{x}(t)$ emanated from slightly different initial conditions $\vec{x}_{0},~\vec{x}_{0}+\vec{\delta}_{0}$. During the time evolution the distance between phase trajectories $\|\vec{\delta}(t)\|=\|\vec{\delta}(0)\|\exp\big(\lambda t\big)$ increases provided that the Lyapunov exponent is positive $\lambda>0$. Therefore, arbitrary small uncertainties in the initial conditions accumulate over a long period to the substantial error, calling thus for a statistical description when evaluating observable quantities.

\subsection{Model of the minimal D=3/2 Hamiltonian chaos}
The kicked rotator is a prototype of systems exhibiting chaos. The Hamiltonian of the one dimensional classical periodically kicked rotator reads
\begin{eqnarray}
    \label{classical rotator}
     && H\big(t\big) = \frac{p}{2}+k\cos\theta\sum\limits_{n}\delta \big(t-n\big).
\end{eqnarray}
Here $(p,\theta)$ is the canonical pair of momentum-angle variables and $n$ specifies the number of  kicks. Solution of the classical kicked rotator problem is described by the Chirikov's map
 \begin{eqnarray}
    \label{Chirikovmap}
     && \theta_{n+1}=\theta_{n}+p_{n+1}, \nonumber \\
     &&  p_{n+1}=p_{n}+k\sin\theta_{n+1}.
\end{eqnarray}
\begin{figure}[ht]
    \centering
    \includegraphics[width=0.9\columnwidth]{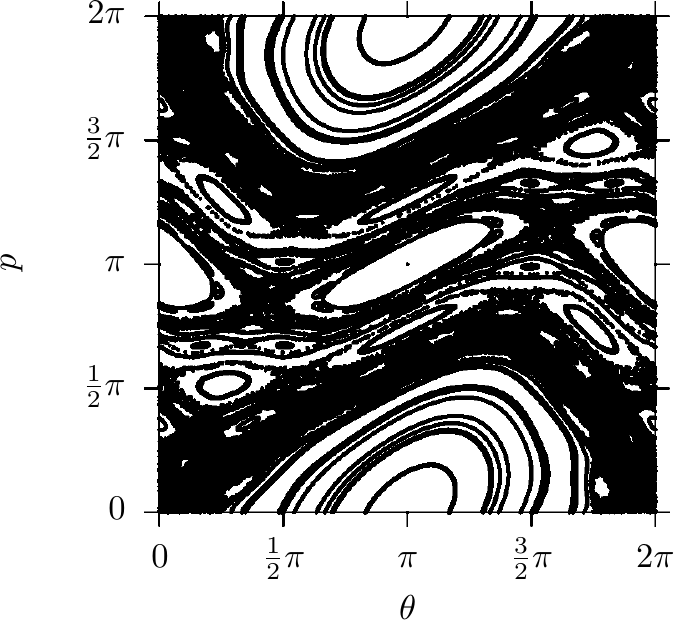}
    \caption{The Chirikov's map with $k=1.0$. The graph was calculated using Julia programming language\cite{Julia} with the \textit{DynamicalSystems.jl} package.}
    \label{figstdmap}
\end{figure}

The phase space of the Chirikov's map consists of regions of regular and chaotic motions. If the strength of the kick exceeds the specific value $k>0.97$,  regular islands disappear, and the chaotic sea covers the whole phase space see Fig.(\ref{figstdmap}) leading to a diffusive growth of the momentum in time (in the number of kicks) $\langle \big(p_{n}-p_{0}\big)^{2}\rangle=\frac{k^{2}}{2}n$.
The kicked rotator is realized in a number of different physical setting. For example, in Ref.\cite{Komnik} the model was employed to investigate spin systems.
Contrary to the classical case, the quantum counterpart of the model Eq.(\ref{classical rotator})
\begin{eqnarray}
    \label{quantum rotator}
     && \hat{H}\big(t\big) = -\frac{1}{2}\frac{\partial^{2}}{\partial \theta^{2}}+k\cos\theta\sum\limits_{n}\delta \big(t-n\big),
\end{eqnarray}
shows a  suppression of  the diffusion due to the destructive role of  quantum coherence \cite{Haake}. Hence, instead of the diffusive growth  localization occurs. We note, in both the classical and the quantum case the
 key issue is the discrete  time evolution. The quantum kicked rotator can be mathematically mapped onto the one dimensional Anderson tight-binding model \cite{Anderson};
\begin{eqnarray}
    \label{Anderson model}
     && T_{m}U_{m}+\sum\limits_{r\neq 0}W_{r}U_{m+r}=EU_{m}.
\end{eqnarray}
Here $W_{r}$ are the hopping amplitudes, the on-site energies $T_{m}$ are uniformly distributed in the interval $T_{m}\in [-T_{max}/2, T_{max}/2]$ and mimic the effect of  randomly embedded impurities. Localization is indicated by the absence of  mobility of electrons in the localized phase and the exponential decay of  amplitudes of the wave functions. If the electron is initially  localized  around  the site $n_{in}$, the probability to find the electron on the site $m$ exponentially decays $U_{m}^{(n_{in})}\approx \exp\big(-|n_{in}-m|/l_{c}\big)$ with the distance $|n_{in}-m|$, and the localization length is a function of energy $l_{c}\big(E\big)$.
The Floquet operator $\hat{F}=\exp(-ik \hat{V}(\theta))\exp(-i\hat{p}^{2}/2)$ describes the evolution of the kicked rotator during the one period of driving. The unperturbed Hamiltonian $\hat{H}_{0}=\hat{p}^{2}/2$ is diagonal $\hat{p}|n\rangle=n|n\rangle$ in the eigenbasis $|n\rangle=\frac{1}{\sqrt{2\pi}}\exp(in\theta)$ of the momentum operator. Therefore, for the wave functions before and after the n-th kick we use the following ansatz
\begin{eqnarray}
    \label{wavefunctionansatz}
     && |\psi^{+}(t+1)\rangle =\hat{F}|\psi^{+}(t)\rangle,\nonumber \\
     && |\psi^{-}(t)\rangle =\exp(-i\hat{H}_{0}/\hbar)|\psi^{+}(t-1)\rangle .
\end{eqnarray}
Here $|\psi^{+}(t+1)\rangle$ and $|\psi^{+}(t)\rangle$ are wave functions after the $n+1$-th and $n$-th kicks, while $|\psi^{-}(t)\rangle$ is the wave function before the $n$-th kick.
For treating the free and kicked evolution parts, we utilize the following parametrization
\begin{eqnarray}
    \label{parametrization}
     && |\psi^{\pm}(t)\rangle =\sum\limits_{n}\psi^{\pm}_{n}(t)|n\rangle,\nonumber \\
      && |\psi^{\pm}(t)\rangle =\int\limits_{0}^{2\pi}d\theta\psi^{\pm}(\theta,t)|\theta\rangle,
\end{eqnarray}
leading to the following recurrent relations for the expansion coefficients:
\begin{eqnarray}
    \label{expansion coefficients}
     &&\psi^{-}_{n}(t+1)=\exp(-in^{2}/2)\psi^{+}_{n}(t),\nonumber \\
     && \psi^{+}(\theta,t)=\exp(-ik\hat{V}(\theta))\psi^{-}(\theta,t).
\end{eqnarray}
Taking into account Eq.(\ref{wavefunctionansatz}), Eq.(\ref{parametrization}) and Eq.(\ref{expansion coefficients}) we deduce the quantum map for the expansion amplitudes in
Eq.(\ref{parametrization})
\begin{eqnarray}
    \label{quantum map}
     && \psi^{+}_{m}(t+1)=\langle m|\psi^{+}(t+1)\rangle=\nonumber\\ &&\langle m|\exp(-ik\hat{V}(\theta))\psi^{-}(t+1)\rangle=\nonumber\\
     &&\sum\limits_{n}\langle m|\exp(-ikV(\theta))|n\rangle\exp(-in^{2}/2)\psi^{+}_{n}(t).
\end{eqnarray}
In the case $\hat{V}=\cos\theta$, the matrix element $J_{nm}=\langle m|\exp(-ikV(\theta))|n\rangle$ becomes  Bessel functions.
To find the equivalence between the kicked rotator and the Anderson model Eq.(\ref{Anderson model}) we use the eigenfunctions and eigenvalues of the Floquet operator $\hat{F}|u^{+}\rangle=\exp(-i\phi)|u^{+}\rangle$, and consider the following identity $u^{-}(\theta)=\exp(ikV(\theta))u^{+}(\theta)$ in the $\theta$ representation.
Taking into account $\exp(-i\phi)\hat{F}u^{+}(\theta)=\exp(-ikV)\exp(i(\phi-\hat{H}_{0}))u^{+}(\theta)=u^{+}(\theta)$ we obtain $u^{-}(\theta)=\exp(i(\phi-\hat{H}_{0}))u^{+}(\theta)$
or in the eigenbasis of $\hat{H}_{0}$ $u^{-}_{n}=\exp(i(\phi-n^{2}/2))u^{+}_{n}$.
Let us introduce the auxiliary operator $\hat{W}$ defined as follows
$\exp(-ik\hat{V})=\frac{1+i\hat{W}}{1-i\hat{W}}=\tan\big(\frac{k\hat{V}}{2}\big)$
and the vector $|u\rangle=\frac{1}{2}\big(|u^{+}\rangle+|u^{-}\rangle\big)$. Then, with  the identity $u(\theta)=(1+i\hat{W})^{-1}u^{+}(\theta)=(1-i\hat{W})^{-1}u^{-}(\theta)$ we infer the operator equation:
\begin{eqnarray}
    \label{operator equation}
     && [1-i\hat{W}(\theta)]u(\theta)=\exp(i(\phi-\hat{H}_{0}))[1+i\hat{W}(\theta)]u(\theta),
\end{eqnarray}
or
\begin{eqnarray}
    \label{convenient operator equation}
     && i\frac{\exp(i(\phi-\hat{H}_{0}))}{\exp(i(\phi+\hat{H}_{0}))}u(\theta)-\hat{W}(\theta)u(\theta)=0.
\end{eqnarray}
Introducing the notations $T_{m}=i\frac{1-\exp(i(\phi-m^{2}/2))}{1+\exp(i(\phi-m^{2}/2))}=\tan\big((\phi-m^{2}/2)/2\big)$,
$u_{m}\langle m|u\rangle$, $E=-W_{0}$, $W_{r}=\langle m|W|m+r\rangle$ we rewrite Eq.(\ref{convenient operator equation}) in the matrix form
\begin{eqnarray}
    \label{matrix form}
     && T_{m}u_{m}+\sum\limits_{r\neq 0}W_{r}u_{m+r}=Eu_{m}
\end{eqnarray}
which demonstrates the mapping  between  the kicked rotator Eq.(\ref{quantum rotator}) and the Anderson tight-binding model
Eq.(\ref{Anderson model}).
%
%
We note that the Anderson model, as well as the kicked rotator, describe single particle noninteracting systems. These models do not capture Many-body localization in interacting systems.
\subsection{D>2 classical and quantum chaos, KAM theorem and random matrix theory}
For nonlinear systems with a few degrees of freedom
 the coupling between different variables is decisive. At certain values of the coupling strength, the
internal resonances overlap giving rise to  Hamiltonian chaos and new phenomena such as Arnold's diffusion in phase space \cite{Arnold}. Let us introduce the canonical pair of action-angle variables:
\begin{eqnarray}
    \label{action-angle variables}
     && I=\frac{1}{2\pi}\oint pdq,~~~\theta=\frac{\partial S(q,I)}{\partial I},
\end{eqnarray}
and present  theHamiltonian of the system in the form:
\begin{eqnarray}
    \label{KAM Hamiltonian}
     && H\big(I,\theta\big)=H_{0}\big(I\big)+\varepsilon V\big(I,\theta\big).
\end{eqnarray}
Here $H_{0}\big(I,\big),~~\forall~I_{i}\equiv I, i\in(1,..N)$ is the integrable part of the Hamiltonian
and $V\big(I,\theta\big),~~\forall~\theta_{i}\equiv I, i\in(1,..N)$ is the perturbation. We presume that the Hamiltonian is not degenerate i.e. the Hessian $det|\frac{\partial \omega}{\partial I_{i}}|=det|\frac{\partial^{2}H_{0}}{\partial I_{i}\partial I_{j}}|\neq0$ is not zero. \\
Following the Kolmogorov–Arnold–Moser (KAM) theorem one finds  that the motion of an integrable system is confined to an invariant torus. If the system is subjected to a weak nonlinear perturbation, some of the invariant tori are deformed and survive.\\
The mechanism of Arnold's diffusion is as follows: In the region of the destroyed invariant torus, in the vicinity of the internal resonances condition $\sum\limits_{j=1}^{N}n_{j}\omega_{j}=0,~~\forall~ n_{j}\in \mathbb{N}$ a homocyclic structure emerges. The internal resonance condition determines a family of surfaces $H_{0}\big(I_{R}\big)=E$ with corresponding resonant action variables $I_{R}$.
Nodal crossing points of different resonances construct a stochastic net. The Arnold's diffusion sets in the stochastic net and is a topological generalization of classical kicked rotator problem.
The Arnold's diffusion was explored for the Heisenberg model, e.g., in Ref.\cite{Chotorlishvili1}.
The specific energy level statistics characterize the quantum counterpart of the classically chaotic nonlinear $D>2$ system Eq.(\ref{KAM Hamiltonian}). Depending on the symmetry of the Hamiltonian $\hat{H}$, in the region of destroyed phase space the invariant tori, according to the random matrix theory (RMT)\cite{Haake} the mean inter-level energy distance $S=\langle E_{n+1}-E_{n}\rangle$ follows either a Gaussian orthogonal ensemble (GOE), a Gaussian unitary ensemble (GUE)  or a Gaussian symplectic ensemble (GUE), respectively, i.e.
\begin{eqnarray}
    \label{KAMstatistics}
     && P(S)=\frac{\pi S}{2}\exp\bigg(-\frac{\pi}{4}S^{2}\bigg),\nonumber\\
     && P(S)=\frac{32 S^{2}}{\pi^{2}}\exp\bigg(-\frac{4 S^{2}}{\pi}\bigg),\nonumber\\
     && P(S)=\frac{2^{18}}{3^{6}\pi^{3}}S^{4}\exp\bigg(-\frac{64}{9\pi}S^{2}\bigg),
\end{eqnarray}
while for the integrable case the statistics is Poissonian $P(S)=\exp(-S)$.
We note that the termination of the quantum Arnold's diffusion for the particle moving in a quasi-1D waveguide was studied in \cite{Demikhovskii}.   The case $D=3/2$  was addressed in Ref. \cite{Chotorlishvili2}.
 Classical chaotic Hamiltonian systems are insensitive to GOE, GUE, and GSE symmetries.
\section{Diagnostic tools of the localized state}
Before analyzing MBL in more details, we briefly review diagnostic tools of a localized state.
\subsection{Multifractality of the wave functions and scaling properties}
\textbf{a} The inverse participation ratio averages the fourth power of the wave function. {It is positive for localized states and vanishes for extended states in the
thermodynamic limit}. If the wave function $|\psi_{i}\rangle$ at site $i$ of a tight-binding model is normalized then partition ratio $\hat{p}_{i}$ (PR) is given by
\begin{eqnarray}
    \label{participation ratio}
     && \hat{p}_{i}=\frac{1}{\bigg(N\sum\limits_{r}|\psi_{i}(r)|^{4}\bigg)}.
\end{eqnarray}
The inverse participation ratio (IPR) has been defined as
\begin{eqnarray}
    \label{inverse participation ratio}
     && \hat{P}_{i}=\sum\limits_{r}|\psi_{i}(r)|^{4}.
\end{eqnarray}
If the wave function spreads over $l$ lattice sites with equal amplitude $|\psi_{i}(r)|^{2}=1/l$
and vanishes elsewhere one can deduce $\hat{p}_{i}=l/N,~~\hat{P}_{i}=1/l$.
As stated by F. Wegner\cite{Wegner},
 {PR describes the proportion of the total number of atoms
in a system which contribute effectively to an eigenstate, whereas IPR is
the inverse number of orbitals contributing effectively to this state. For
localized states, IPR is larger than zero whereas PR vanishes in the
thermodynamic limit}.
The wavefunction of the system shows anomalous scaling properties in the vicinity of the transition point to the localized phase. In particular, the ensemble averaged quantity \cite{Evers}
\begin{eqnarray}
    \label{fractal properties}
     && P^{(k)}(E)=\bigg\langle\sum\limits_{i,r}|\psi_{i}(r)|^{2k}\delta(E-e_{i})/
     \sum\limits_{i}\delta(E-e_{i})\bigg\rangle, \nonumber\\
     && P^{(k)}\approx L^{-\tau_{k}},~~\tau_{k}=d(k-1)+\Delta_{k}.
\end{eqnarray}
shows fractal scaling properties with the characteristic size of the system $L$ and non-integer $\tau_{k}$.
Here $\psi_{i}(r)$ is the amplitude of the eigenfunction $|i\rangle$ with the energy $e_{i}$ at site $r$ of the particle in a
tight binding model and non-integer $\Delta_{k}\notin \mathbb{N}$.
\subsection{RMT beyond Quantum chaos. The transfer-matrix method\cite{Beenakker}}
Let us consider the simplest formulation of the scattering problem, a scattering of an electron on the $\delta$-like impurities located at $x = x_{0}$.
The Schr\"odinger equation
\begin{eqnarray}
    \label{scattering problem}
     && i\frac{\partial \psi}{\partial t}=-\frac{1}{2m}\bigg(\frac{\partial^{2}\psi}{\partial x^{2}}+\Lambda \delta (x-x_{0})\psi\bigg)
\end{eqnarray}
admits the solution
\begin{eqnarray}
    \label{scattering problem solution}
     && \psi(x)=a^{+}\exp(ikx)+a^{-}\exp(-ikx),~~~x<x_{0},\nonumber\\
     && \psi(x)=b^{+}\exp(ikx)+b^{-}\exp(-ikx),~~~x>x_{0}.
\end{eqnarray}
and the continuity conditions
\begin{eqnarray}
    \label{continuity conditions}
     && \psi(x^{+}_{0})=\psi(x^{-}_{0}),\nonumber\\
     && \partial_{x}\psi(x^{+}_{0})-\partial_{x}\psi(x^{-}_{0})=\Lambda \psi(x_{0}).
\end{eqnarray}
define the elements of the scattering matrix:
\(
c_{out}=\begin{bmatrix}
b^{+}\\
b^{-}
\end{bmatrix}
\)
=
\(
\begin{bmatrix}
r & \acute{t}\\
t &  \acute{r}
\end{bmatrix}
\)
\(
\begin{bmatrix}
a^{+}\\
a^{-}
\end{bmatrix}
\)
=
\(
S\begin{bmatrix}
a^{+}\\
b^{-}
\end{bmatrix}
=Sc_{in}.\)\\
Generally, the channel can be arbitrarily large and the dimension of the $t$ matrix is equal to the number of  scattering channels.
The explicit expressions of the $t$ matrix elements is obtained  after solving  Eq.(\ref{continuity conditions}) and expressing the  coefficients $b^{+},b^{-}$
in terms of $a^{+},a^{-}$. The Landauer-B\"uttiker conductance\cite{Beenakker} is given by:
\begin{eqnarray}
    \label{Landauer}
     && G=\frac{2e^{2}}{\hbar}\sum\limits_{n}T_{n}=\frac{2e^{2}}{\hbar}Tr\big(t^{\dag}t\big).
\end{eqnarray}\\
We note, {a)} {When the system is in the localized phase, the eigenvalues of the transfer matrix are random numbers. However, there is a principle difference between RMT of quantum transport and chaotic nonlinear systems. The RMT describes quantum transport in terms of transmission eigenvalues of the open system (i.e., leads, impurities, etc.), while RMT is an  internal property of nonlinear non-integrable quantum systems.}
The Thouless energy characterizes the energy scale of disordered conductors and is defined via the formula $E_{T}=\hbar D/L^{2}$, where $D$ is the diffusion constant, and $L$ is the characteristic size of the system.
{b)} {B. L. Altshuler and B. I. Shklovskii \cite{Altshuler,eeanderson} showed that for energy separations greater than the Thouless energy $|E^{'}-E|>E_{T}$, the correlation function deviates from random-matrix theory.}
If the condition $|E^{'}-E|<E_{T}$ holds,   the distribution function of eigenvalues of the transfer matrix read
\begin{eqnarray}
    \label{RMTtransfermatrix}
     && P(S)=c \exp[-\beta Tr f(t^{\dag} t)],\nonumber \\
     && P\big(\{T_{n}\}\big)=c\prod\limits_{i<j}|T_{i}-T_{j}|^{\beta}\times\nonumber\\
     && \prod\limits_{k}T_{k}^{-1+\beta/2}\exp[-\beta f(T_{k})].
\end{eqnarray}
Here $P\big(\{T_{n}\}\big)$ describes  the correlated distribution of eigenvalues,  $S=\langle E_{n+1}-E_{n}\rangle$ is the mean distance between neighboring  eigenvalues, $\beta=1$ for GOE, $\beta=2$ for GUE, and $\beta=4$ for GSE symmetries, respectively, and the function $f(T_{k})$ is determined through the average spectral density $f(T_{k})=\int dT^{'}_{k}\sigma(T^{'}_{k})\ln|T_{k}-T^{'}_{k}|$.
\section{Energy level statistics for the MBL phase}
MBL is a localization phenomenon in  interacting system
\cite{Basko,Devakul,Luitz,Vasseur,Bardarson,Serbyn,topical,huse,huse1,huse2,Bauer}, distinct from the Anderson localization:
We note, in  absence of  electron-electron interaction in low dimensional systems $d=1,2$ all states are localized even in the presence of arbitrary small disorder and the conductivity $\sigma_{d=1,2}(T)=0$. For $d=3$ the conductivity follows the Arrhenius law $\sigma_{d=3}(T)\sim \exp(-E_{c}/T)$. Here $E_{c}$ is the distance between the Fermi level and the mobility edge (mobility edge refers to the border between localized and extended bands). Therefore, for any finite $E_{c}$, the Anderson localization of electronic states leads to a metal-insulator transition at zero temperature. The conductivity becomes finite at any finite temperature. MBL can be viewed as Anderson like localization of many-body wave functions.
The expectation value of an arbitrary quantum operator is given by formula $\langle\mathcal{\hat{O}}\rangle_{t}=\langle \psi(t)|\mathcal{\hat{O}}|\psi(t)\rangle$. Taking into account that $|\psi(t)\rangle=\exp(-i\hat{H}t/\hbar)|\psi(0)\rangle$ and utilizing the expansion over the basis of energy eigenstates $|\psi(0)\rangle=\sum\limits_{\alpha}c_{\alpha}|E_{\alpha}\rangle$ we deduce: $\langle\mathcal{\hat{O}}\rangle_{t}=\sum\limits_{\alpha,\beta}c_{\alpha}^{\ast}c_{\beta}\mathcal{O}_{\alpha\beta}\exp(-i(E_{\beta}-E_{\alpha})t/\hbar)$, where $\mathcal{O}_{\alpha\beta}=\langle E_{\alpha}|\mathcal{\hat{O}}|E_{\beta}\rangle$.
{The Eigenstate Thermalization Hypothesis(ETH) states\cite{Ilievski,Pozsgay,Mierzejewski,ETH1,ETH2,ETH3} that for quantum ergodic systems, in the long time limit the expectation value is given by the following formula:  $\langle\mathcal{\hat{O}}\rangle_{t}^{t\rightarrow\infty}=\sum\limits_{\alpha}|c_{\alpha}|^{2}\mathcal{O}_{\alpha\alpha}$.
The MBL state violates ETH. Therefore, MBL can be referred to as a non-ergodic phase. In the ergodic phase, depending on the symmetry of the Hamiltonian $\hat{H}$, the  energy level statistics is either GOE, GUE, or GSE, while in the MBL phase level statistics is Poissonian.}
The transition from the ergodic phase to the MBL phase occurs at specific critical strength of disorder. To infer precisely the MBL transition point, one needs to obtain and analyze a large number of statistics, i.e., consider different realizations of disorder and take ensemble average. A key quantity is the disorder average of the ratio
\begin{eqnarray}
    \label{statistics}
     && r_n = \min \left( \delta_n, \delta_{n-1} \right) / \max \left(\delta_n, \delta_{n-1}\right),\nonumber\\
     && r = \frac{1}{N-2} \sum\limits_{n=3}^{N} r_n
\end{eqnarray}
where $\delta_n = E_n - E_{n-1}$ is the distance between two neighboring energy levels {labelled by $n$} and $N$ is the number of eigenstates.
In the ergodic phase for the Gaussian orthogonal ensemble GOE $r_\texttt{{GOE}} = 0.5307$ is found, while in the MBL phase $r_\texttt{{Poisson}} = 0.3863$. The disorder strength has a strong influence on the system's spectral characteristics. To avoid finite size effects and other numerical artifacts calculations should be done for different sizes of the  system and then the obtained data
should be collapsed into a single universal size-independent result.  The conventional finite-size scaling analysis works as follows\cite{Luitz}:
One selects the middle part of the spectrum of the system. The level statistics of $\left\langle r \right\rangle$ as a function of disorder $h$ are scaled with $f_L(h) = L^{1/\nu} (h-h_c)$, where $\nu = 1$ was assumed and $h_c$ defines the critical strength of disorder. For best fitting parameter $h_c$, the scaling procedure  admits   a minimizing function $w(h)$
\begin{equation}
     w = \sum\limits_{L, L^\prime} \int\limits_{h_1}^{h_2} \left| \left\langle r \right\rangle(f_L(h)) - \left\langle r\right\rangle(f_{L^\prime}(h))  \right| d h
     \label{eq:min}
\end{equation}
where $h_1$ and $h_2$ are defined by the common integration domain
$    h_1 = \max \left( L_i^{\frac{1}{\nu}} (a - h_c) \right),\;
    h_2 = \min \left( L_i^{\frac{1}{\nu}} (b - h_c) \right)$
where $L_i$ denote the system sizes to be analyzed and $a$, $b$ are boundaries limits for un-scaled data set.
\section{Systems with complex symmetry properties}
	For systems with a mixed symmetry, $\hat{H}_{tot}=\hat{H}_{GOE}+\lambda\hat{H}_{GUE}$ the total Hamiltonian doesn't belong to a particular symmetry class. In this case, finite-size scaling analysis may fail when GOE/GUE statistics show different scaling features. Depending on the value of the parameter $\lambda$ the spectrum of the Hamiltonian $\hat{H}_{tot}=\hat{H}_{GOE}+\lambda\hat{H}_{GUE}$ displays different features: In the limit of small $\lambda$  the  level statistics obey GOE, for large $\lambda$  GUE, and what is interesting, in the transition area the system shows qualitatively different properties than $\hat{H}_{GOE}$ and $\hat{H}_{GUE}$. We need a new method to explore the MBL phase in systems with mixed complex symmetries.

\begin{figure}[!htb]
    \centering
    \includegraphics[width=0.98\columnwidth]{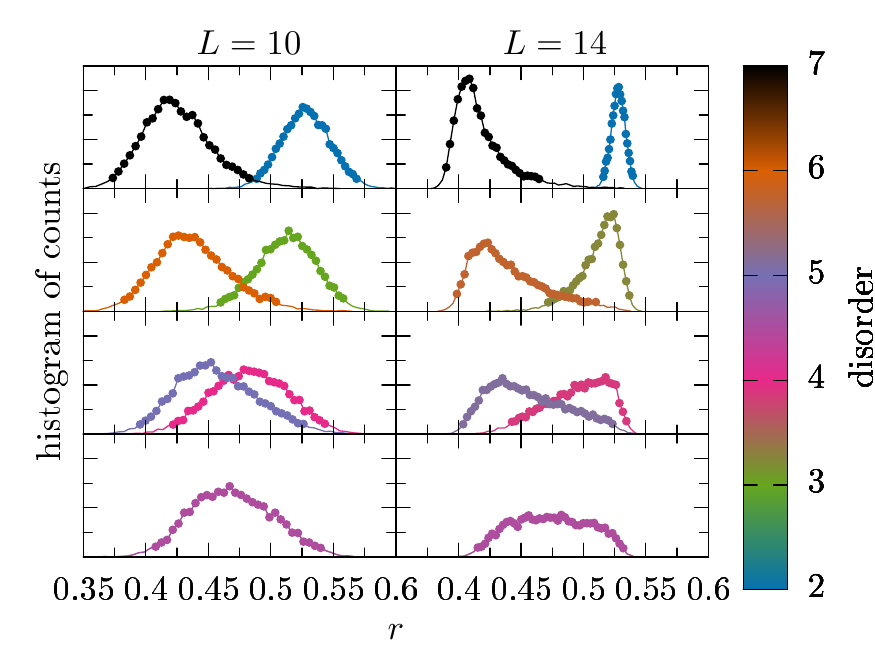}
    \caption{
        Histograms of counts for the different strength of disorder $h$ with $J_1 = -1$ without DMI as a function of the consecutive level spacing $r$. Broadening of the histogram corresponds to the critical strength of disorder and to the transition point. The larger is $L$ the peaks are more distinguished.
    }
    \label{fig:brd}
\end{figure}
\begin{figure}[!htb]
    \centering
    \includegraphics[width=0.99\columnwidth]{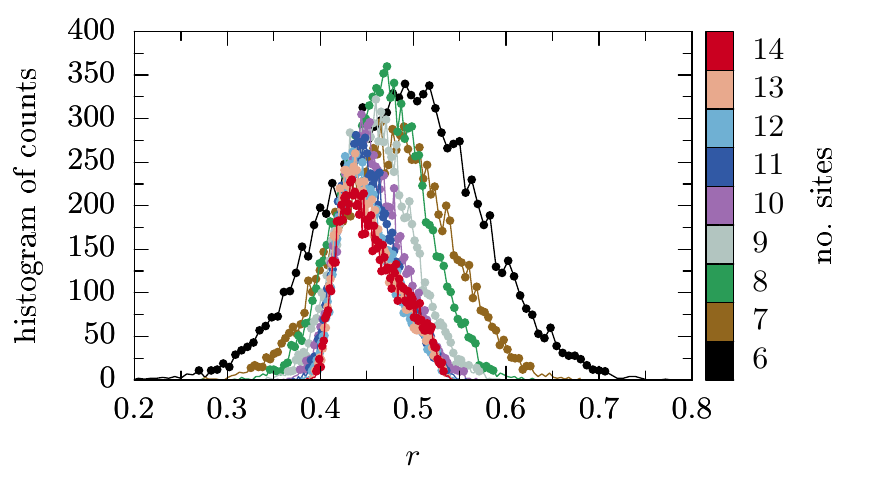}
    \caption{
        Histograms of counts for a fixed strength of disorder $h = 5$ without DMI as a function of the consecutive level spacing $r$. Convergence is indicated  for L=14.
    }
    \label{fig:brd_L}
\end{figure}

\begin{figure}[ht]
    \centering
    \includegraphics[width=0.7\columnwidth]{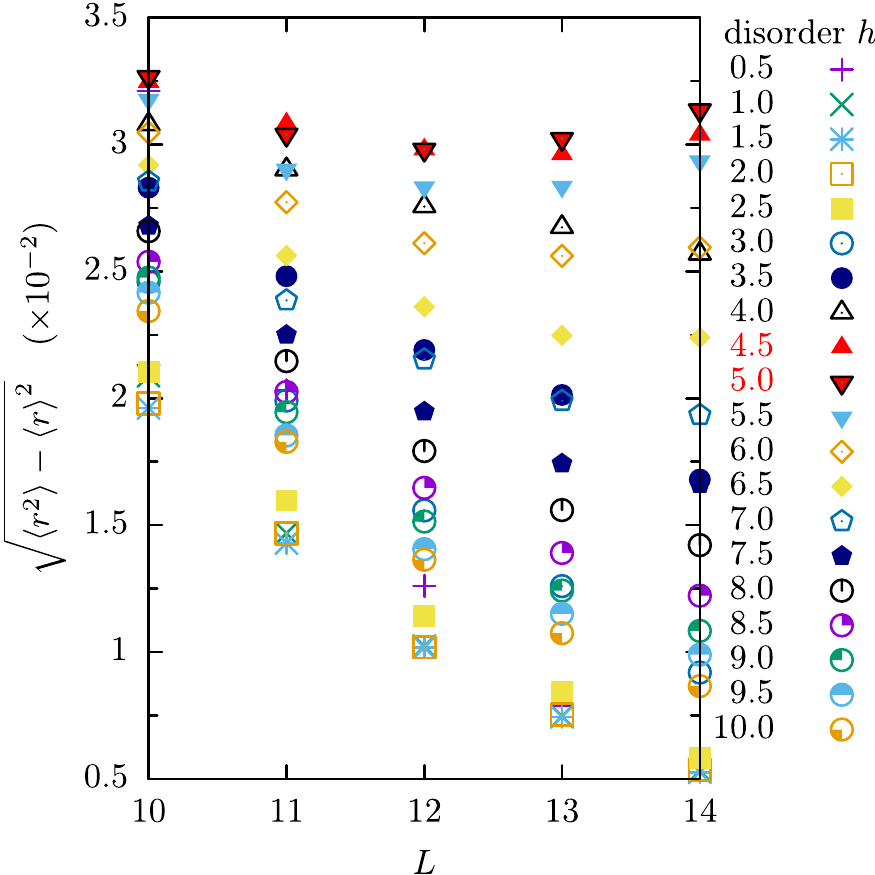}
    \caption{
        The fluctuation dependency on system size without $D$ and $J_1 = -1$. The two datasets marked by red are the nearest to critical disorder.
    }
    \label{fig:fluct}
\end{figure}
\begin{figure}[ht]
    \centering
    \includegraphics[width=0.98\columnwidth]{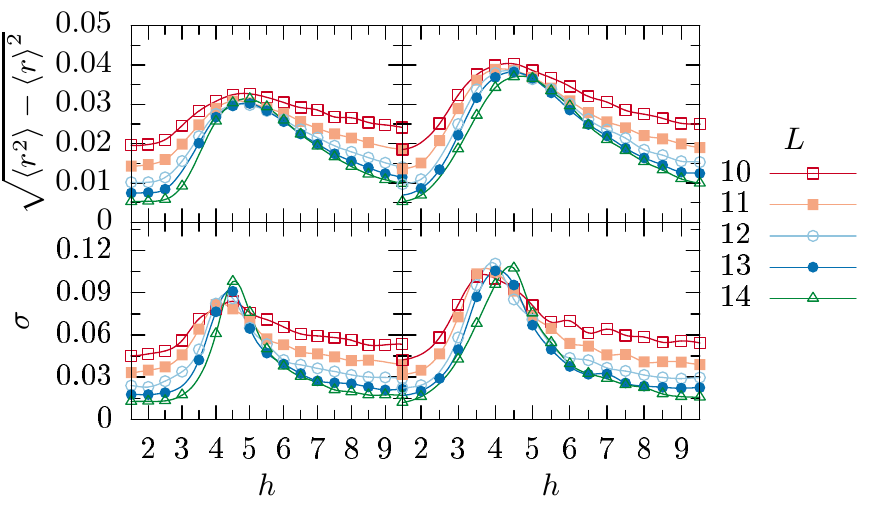}
    \caption{ Full width at half maximum $\sigma$ for the histograms  as a function of disorder $h$. The graphs on right side are with finite DMI, $D = 0.2$.}
    \label{fig:fss}
\end{figure}
\begin{figure}[ht]
    \centering
    \includegraphics[width=0.8\columnwidth]{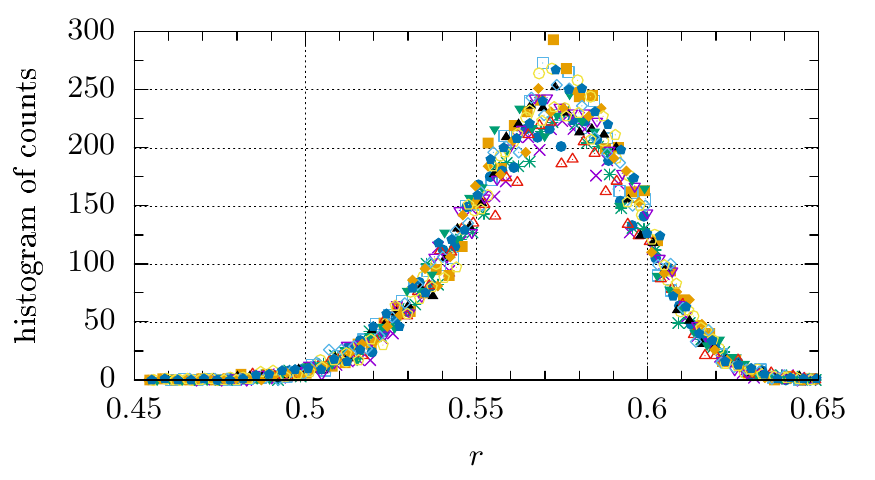}
    \caption{$J_1 = -1$, $D=0.2$, for fixed $h=5$. Additional site-dependent disorder of $10$\% in $D$ was taken while different colors and marks refer to various realizations.}
    \label{fig:d}
\end{figure}

In our recent work \cite{stefan} we observed that in spite of the difference of the GOE and GUE statistics, the enhanced fluctuations at the MBL transition point bears universal physical character typical for the both GOE and GUE symmetry. Thus, our method can be implemented for systems with complex symmetry properties when standard finite-size scaling procedure fails.
We explored the MBL problem Hamiltonian with a dynamical Dzyaloshinskii Moriya interaction~(DMI)~\cite{chiral, Azimi}
\begin{align}
\label{Hamiltonian}
  \hat{H} &= J_1 \sum^{L}_{i=1}\hat{\vec{S}}_i \cdot \hat{\vec{S}}_{i+1} + J_2 \sum^L_{i=1}\hat{\vec{S}}_i\cdot \hat{\vec{S}}_{i+2}\nonumber\\
   &\quad+\sum^L_{i=1} B^z_i \hat{S}_i^z + D \sum^{L}_{i=1} \left(\hat{\vec{S}}_{i} \times \hat{\vec{S}}_{i+1} \right)_z,\; D=E_y g_{ME}.
\end{align}
Note, in the absence of the DMI term  the Hamiltonian has GOE symmetry and the symmetry of the DMI term is GUE. The DMI constant $D$ combines the effect of an electric field and the magnetoelectric coupling. The nearest neighbor exchange interaction is ferromagnetic $J_1<0$ while the next nearest neighbor one is antiferromagnetic $J_2>0$ leading in general to a frustrated spin order.

Computationally we are able to deal with only small chains; experimentally relevant are for instance the Fe chains on the $(5\times 1)-$Ir(001) surface \cite{Wiesendanger}.
$\hat H$ is block-diagonal. Each block is identified via the conserved total spin component $\hat{S}^{z}=\sum^L_{i=1}\hat{S}_i^{z}$. Of a special interest is the largest subspace of states $|\Psi_n\rangle$ { obeying $\hat{S}^{z} |\Psi_n\rangle=M |\Psi_n\rangle$ with $M=0$ for even $L$ or $M=1$ for odd $L$, respectively}. A uniform magnetic field $B^z_i = B^z$ shifts the eigenvalues equally in each subspace and has no prominent effect on the interlevel distance $r_n$, while randomness incorporated in the magnetic field $B^z_i \in \left\langle -h, h \right\rangle$ can induce a qualitative change of the spectral properties from Wigner-Dyson to Poisson level spacing statistics. The strength of disorder is measured on a scale set by $J_{1}$. In what follows we work with dimensionless units such that $J_{1}=1$.

 To formulate a possibly general criterion for MBL that is applicable in such cases, as well we analyze the full statistics for each realization $\alpha$ of the random magnetic fields $r^{(\alpha)}$. The histograms corresponding to a counting classification of $r^{(\alpha)}$ for a given disorder strength $h$ is presented on Fig.~\ref{fig:brd}. As can be inferred, the histograms are narrow far away from the MBL transition, while the histograms become particularly broad close to the transition point mimicking the behavior of fluctuations near conventional phase transitions. The histograms become more and more pronounced for increasing $L$.
Fig.~\ref{fig:brd_L} demonstrates the convergence of the histograms of for chains with different lengths. As we see already for $L=14$ counts histograms amalgamate underlying that an analysis of the histograms can serve as a further indicator in addition to finite-size scaling. The convergence of histograms even for relatively small systems endorses our method as less computationally demanding which is a major advantage for exact diagonalization approaches
that are considered as well suited for MBL studies.
As for the histograms of consecutive level spacing, Fig.~\ref{fig:brd} illustrates the broadening of the histograms when approaching the transition point between the ergodic and the MBL phases. As evident, the effect of broadening is even more prominent for systems with a larger size Fig.~\ref{fig:brd}. We note that the observed phenomena is not related to a particular type of level statistics but it is rather akin to the transition regime. Away from the transition point on the ergodic side (GOE statistics), and on the MBL phase side (Poisson statistics) the width of histograms are narrower. The broadening is linked to the enhanced quantum fluctuations Fig.~\ref{fig:fluct}.
This behavior is of a general character and is maintained even after adding the next nearest neighbor interaction and DMI terms.

Physically, the broadening of  histograms  is attributable to the enhanced fluctuations near phase transitions (cf.~Figs.~\ref{fig:fluct}, \ref{fig:fss}). Hence, such broadening serves as a further indicator for approaching the MBL phase.
%

%

%
Disorder in the exchange coupling or in $D$ may also  occur. The latter (cf.~Eq.~\ref{Hamiltonian}) can be  viewed as
 random change in $ \mathbf E$ or a random elastic energy change
($ \mathbf E.{\bf P} =g_{ME} \mathbf E \sum\limits^L_{i=1} \langle \mathbf{e}_x \times
  ( \mathbf{\hat S}_i \times \mathbf{\hat S}_{i+1} )\rangle $),
 and thus, it is important for spin-phonon-coupled systems at finite temperatures. Calculations evidence the robustness of the
MBL phase against randomizing $D$ within a physically reasonable range, an example is depicted in  Fig.~\ref{fig:d}.\\
\acknowledgments
We are indebted to Marcin Mierzejewski for numerous discussions and suggestions.
We acknowledge financial support from the Deutsche Forschungsgemeinschaft through
SFB 762, and BE 2161/5-1.


\newpage
\begin{thebibliography}{99}
\bibitem{Lorenz} E. Lorenz, J. Atmos. Sci. \textbf{20}, 130 (1953).
\bibitem{Zaslavsky} G. M. Zaslavsky \textit{Physics of Chaos in Hamiltonian Dynamics}, (London Imperial College Press, 1998).
\bibitem{Komnik}
L. Chotorlishvili, Z. Toklikishvili, A. Komnik, and J. Berakdar, Phys. Rev. B \textbf{83}, 184405 (2011).
\bibitem{Haake}  F. Haake, \textit {Quantum Signatures of Chaos}, (Springer, Berlin, 2000);
\bibitem{Fishman}  S. Fishman, D. R. Grempel, and R. E. Prange Phys. Rev. Lett. \textbf{49}, 509 (1982).
\bibitem{Anderson} P. W. Anderson Phys. Rev. \textbf{109}, 1492 (1958).
\bibitem{Arnold} V. I. Arnold \textit{Mathematical Methods of Classical Mechanics}, (Springer-Verlag, New York 1989).
\bibitem{Chotorlishvili1} L. Chotorlishvili, Z. Toklikishvili, and J. Berakdar, J. Phys. Condens. Matter \textbf{21}, 356001 (2009).
\bibitem{Demikhovskii} V. Ya. Demikhovskii, F.M. Izrailevb, A. I. Malyshev,  Phys. Lett. A \textbf{352}, 491 (2006).
\bibitem{Chotorlishvili2} L. Chotorlishvili, A. Ugulava, Physica D \textbf{239}, 103 (2010).
\bibitem{Wegner} F. Wegner Z. Physik B \textbf{36}, 209 (1980).
\bibitem{Evers} F. Evers and A. D. Mirlin Rev. Mod. Phys. \textbf{80}, 1355 (2008).
\bibitem{Beenakker} C. W. J. Beenakker Rev. Mod. Phys. \textbf{69}, 731 (1997).
\bibitem{Altshuler} B. L. Altshuler and B. I. Shklovskii, Zh. Ekso. Teor. Fiz. \textbf{91}, 220 (1986).
\bibitem{eeanderson} \textit{Electron-electron Interactions in Disordered Systems},. Eds. M.Pollak and A.L. Efros (North-Holland, Amsterdam, 1984).
\bibitem{Basko} D.M. Basko, I. L. Aleiner, B. L. Altshuler Annals of Physics \textbf{321}, 1126 (2006).
\bibitem{Devakul} L. Zhang, B. Zhao, T. Devakul, D. A. Huse, Phys. Rev. B \textbf{93}, 224201 (2016).
\bibitem{Luitz}  D. J. Luitz, N. Laflorencie, and F. Alet, Phys. Rev. B \textbf{91}, 081103(R) (2015); D. J. Luitz, N. Laflorencie, and F. Alet, Phys. Rev. B \textbf{93}, 060201(R) (2016).
\bibitem{Vasseur}  R. Vasseur, A. J. Friedman, S. A. Parameswaran, and A. C. Potter, Phys. Rev. B \textbf{93}, 134207 (2016).
\bibitem{Bardarson} J. A. Kj\"all, J. H. Bardarson, and F. Pollmann, Phys. Rev.Lett. \textbf{113}, 107204 (2014).
\bibitem{Serbyn}  M. Serbyn and J. E. Moore, Phys. Rev. B \textbf{93}, 041424(R) (2016).
\bibitem{topical} For an overview we refer to the topical issue {\it Many-Body Localization}  ed. J. H. Bardarson {\it  et al.} Ann.~Phys.  (2017).
\bibitem{huse}  V. Oganesyan, A. Pal, D. A. Huse, Phys. Rev. B \textbf{80}, 115104 (2009)
\bibitem{huse1} M. Serbyn, Z. Papic, and D. A. Abanin, Phys. Rev. Lett. \textbf{111}, 127201 (2013).
\bibitem{huse2} D. A. Huse and V. Oganesyan, Phys. Rev. B \textbf{90}, 174202 (2014).
\bibitem{Bauer}  B. Bauer and C. Nayak, J. Stat. Mech.: Theory Exp. \textbf{P09005} (2013); V. Oganesyan and D. A. Huse, Phys. Rev. B \textbf{75}, 155111 (2007);
 A. Pal and D. A. Huse, Phys. Rev. B \textbf{82}, 174411 (2010); C. R. Laumann, A. Pal, and A. Scardicchio, Phys. Rev. Lett. \textbf{113}, 200405 (2014).
\bibitem{Ilievski}  E. Ilievski, J. De Nardis, B. Wouters, J.-S. Caux, F. H. L. Essler, and T. Prosen Phys. Rev. Lett. \textbf{115}, 157201 (2015).
\bibitem{Pozsgay}  B. Pozsgay, M. Mestyán, M. A. Werner, M. Kormos, G. Zaránd, and G. Takács Phys. Rev. Lett. \textbf{113}, 117203 (2014).
\bibitem{Mierzejewski}  M. Mierzejewski, Peter Prelovšek, and Tomaž Prosen Phys. Rev. Lett. \textbf{113}, 020602 (2014).
\bibitem{ETH1} J. Eisert, M. Friesdorf and C. Gogolin, Nat. Phys.\textbf{ 11},  124 (2015).
\bibitem{ETH2} C. Gogolin and J. Eisert, Rep. Prog. Phys.\textbf{ 79},  056001 (2016)
\bibitem{ETH3} L. D’Alessio, Y. Kafri, A. Polkovnikov, and M. Rigol, Adv. Phys. \textbf{65}, 239–362 (2016).
\bibitem{stefan} S. Stagraczy\'nski, L. Chotorlishvili, M. Sch\"uler, M. Mierzejewski, J. Berakdar, Phys. Rev. B \textbf{96}, 054440 (2017).
\bibitem{chiral} H. Katsura, N. Nagaosa, and A. V. Balatsky, Phys. Rev. Lett. \textbf{95}, 057205 (2005);
M. Mostovoy, Phys. Rev. Lett. \textbf{96}, 067601 (2006);
I. A. Sergienko and E. Dagotto, Phys. Rev. B \textbf{73}, 094434 (2006);
Y. Tokura, S. Seki, and N. Nagaosa, Rep. Prog. Phys. \textbf{77}, 076501 (2014);
Y. Tokura and S. Seki, Adv. Mater. \textbf{22}, 1554 (2010);
K. F. Wang, J. M. Liu, and Z.F. Ren, Adv. Phys. \textbf{58}, 321 (2009).
\bibitem{Azimi} M. Azimi, L. Chotorlishvili, S. K. Mishra, S. Greschner, T. Vekua, and J. Berakdar, Phys. Rev. B \textbf{89}, 024424 (2014);
M. Azimi, L. Chotorlishvili, S. K. Mishra, S. Greschner, T. Vekua, W.  H\"ubner and J. Berakdar, New J. of Physics \textbf{16}, 063018 (2014);
M. Azimi, M. Sekania, S. M. Mishra, L. Chotorlishvili, Z. Toklikishvili, and J. Berakdar, Phys. Rev. B \textbf{94}, 064423 (2016);
L. Chotorlishvili, M. Azimi, S. Stagraczy\'nski, Z. Toklikishvili, M. Sch\"uler, J. Berakdar, Phys. Rev. E \textbf{94}, 032116 (2016).
\bibitem{Wiesendanger}  M. Menzel, Y. Mokrousov, R. Wieser, J. E. Bickel, E. Vedmedenko, S. Bl\"ugel, S. Heinze, K. von Bergmann, A. Kubetzka, and R. Wiesendanger,
Phys. Rev. Lett. \textbf{108}, 197204 (2012).
\bibitem{Julia} J. Bezanson, A. Edelman, S. Karpinski, and V. B. Shah, SIAM Rev. \textbf{59}, 65 (2017); G. Datseris, Journal of Open Source Software \textbf{3}, 598 (2018).
\end{thebibliography}
\end{document}